\documentclass[prl, aps, amssymb,amsmath,twocolumn,showpacs,floatfix]{revtex4}
\usepackage{epsfig}
\usepackage{color}
\usepackage{graphicx}

\newcommand{\NM}{{\mathbb N}}

\newcommand{\ZM}{{\mathbb Z}}

\begin{document}
\pagecolor{green}
\color[rgb]{0.5,0.7,0.8}
\title{Quantum Accelerator Modes near Higher-Order Resonances.}
\author{Italo Guarneri and  Laura Rebuzzini
} \affiliation {
{\small  Center for Nonlinear and Complex Systems}\\
{\small  Universit\'a dell'Insubria, via Valleggio 11, I-22100 Como, Italy.}\\
{\small  Istituto Nazionale di Fisica Nucleare, Sezione di Pavia,
via Bassi 6, I-27100 Pavia, Italy.} }


\date{\today}


\begin{abstract}
 Quantum Accelerator Modes have been experimentally observed, and theoretically
explained,
in the
dynamics of kicked cold atoms in the presence of gravity, when the
kicking period  is close to a half- integer  multiple of the Talbot
time. We generalize the theory to the case when the kicking period is
sufficiently close to any rational multiple of the Talbot time, and thus predict
new rich families of experimentally observable Quantum Accelerator
Modes.

\end{abstract}


\maketitle

\noindent Present-day experimental techniques
afford almost perfect control of the state and time evolution of
quantum systems, and thus allow observation of phenomena, that are rooted
in subtle aspects of the quantum-classical correspondence.
In particular, effects of mode-locking  and nonlinear resonance, that are ubiquitous
in classical nonlinear dynamics, could be observed on the quantum
level, in the form of  unexpected quantum stabilization
phenomena; for
instance, in nondispersive wave-packet dynamics \cite{MNG05}, and in
the kicked dynamics of cold and ultra-cold atoms. In the latter case,
techniques originally introduced  by M. Raizen and
coworkers  have been successfully used
to produce atom-optical  realizations of the Kicked Rotor (KR) model \cite{KR},
which is
a famous paradigmatic model of Quantum Chaos.
A variant of the KR, which was realized in
Oxford, had the (Cesium)
atoms freely falling under the effect of gravity between kicks. Discovery of a new effect
followed, which was named
Quantum Accelerator Modes (QAM)
\cite{Ox99}. A natural internal time scale for the system is set
by the so-called Talbot time, and whenever  the kicking period is close to a
 half-integer multiple of that time,
small groups  of atoms are observed to steadily
accelerate away from the bulk of the atomic cloud, at a rate and in
a direction (upwards, or downwards) which depend on parameter
values.
 A theory for this phenomenon \cite{FGR03}  introduces a
dimensionless parameter $\epsilon$, which measures the detuning from
exact resonance, and shows that the nearly resonant quantum dynamics
may be obtained from quantization of a certain classical dynamical
system \footnote{In the present paper, "Classical dynamical system"
has the mathematical meaning, of a system that is endowed
with a finite-dimensional phase space, wherein evolution  is
described by deterministic trajectories.}, {\it using $\epsilon$ as
the Planck's constant}.
This dynamical system was termed the
$\epsilon$-classical limit of the quantum dynamics, and is quite
different from the system, which is obtained in the classical
limit proper $\hbar\to 0$.
QAM  are absent in the latter limit,  and are accounted for
by $\epsilon$-classical phase
space structures. Thus, they are at once a purely quantal phenomenon, and
a manifestation of classical nonlinear resonance; indeed, their theory is
a  repertory of classic items of nonlinear dynamics, occurring
in a purely quantum context. For instance, they are  associated
with Arnol'd tongues in the space of parameters, and
are hierarchically organized according to
number-theoretical rules \cite{farey}.
Finally, on the quantum level, a deep relation to the famous problem of Bloch 
oscillations and Wannier-Stark resonances \cite{WS} has been exposed \cite{SFGR06}.  \\
Existence of QAM somehow related to other {\it rational} multiples of
the Talbot time ("higher order resonances"), than just the  half-integer
ones, is a long-standing question, that lies beyond the reach of
the existing theory. Some indications  in this sense
are given by numerical simulations, and also by  generalizations of
heuristic arguments \cite{GS}, which were formerly devised \cite{Ox99}
in order to explain the
first experimental  observations of  QAM.
\\
In this paper we show that QAM indeed exist near resonances of arbitrary order.
This noticeable re-assessment of the QAM phenomenon requires 
a nontrivial reformulation of the small-$\epsilon$ approximation, in order to circumvent
the basic difficulty, that no $\epsilon$-classical
limit exists in the case
of higher resonances. We show that, in spite of that,
families of rays
(in the sense of geometrical optics) nonetheless exist,
that give rise to QAM in the vicinity
of a KR resonance. Such "accelerator  rays" are
not trajectories of a
single formally classical system, but rather come in families,
generated by different classical systems, which provide
but local (in phase space) approximations to the quantum dynamics.
This is remindful of  the small-$\hbar$ asymptotics
for the dynamics of
particles,
in the presence
of spin-orbit interactions
 \cite{LF91}. This similarity is by no means accidental, because the
KR dynamics at higher-order resonance may be  described in
terms of  spinors \cite{IS80,SZAC}; thus, the present problem naturally fits
into a more general theoretical framework,
and our formal approach
may find application in the broader context of quantum kicked dynamics, in the
presence of spin.
\\The dynamics of kicked atoms moving in the vertical direction
under the effect of gravity is modeled by the following
time-dependent Hamiltonian:
\begin{equation}
\label{ham1} {\hat H}(t)\;=\; \frac12({\hat
P}+\frac{\eta}{\tau}t)^2\;+\;k V({\hat
X})\;\sum_{n=-\infty}^{+\infty}\delta(t-n\tau)\;.
\end{equation}
\begin{figure}
\includegraphics[width=8cm,angle=0]{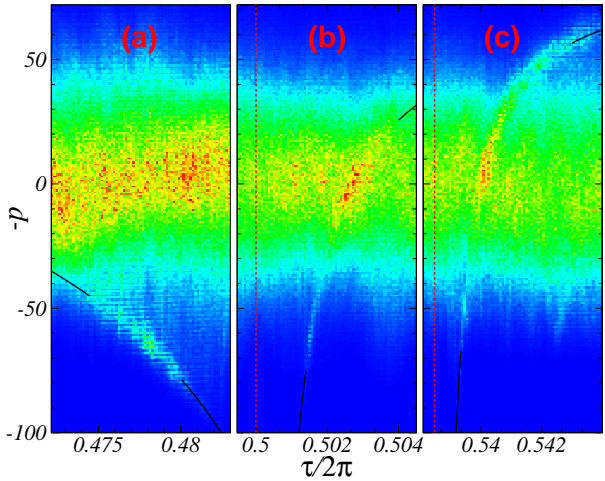}
\caption{(Color online)
Momentum distributions, in the time dependent gauge, after
$n=100$ kicks, for different values of the kicking period near the resonance
$\tau =\pi$ ((a) and (b)), and near the resonance
$\tau =14\pi/13$ (c). Red color corresponds to highest probability.
The initial state is a mixture of 100 plane waves sampled from
a gaussian distribution of momenta. Vertical dashed lines correspond to the
mentioned resonant values
 of $\tau /2\pi$. Black full lines show the theoretical curves (\ref{acce}), with:
(a) $T=2, {\mathfrak p}=3,
{\mathfrak j}=1, \Delta _2 =0$ , (b) $T=1, {\mathfrak p}=5, {\mathfrak j}=1, \Delta _1 =0$
and (c) $T=1, {\mathfrak p}=5, {\mathfrak j}=1, \Delta _1 =20\pi /13$.
Parameter values are:
$k=0.8\pi$ and $\eta =0.126\tau$.
}
\label{pac}
\end{figure}
 Units are chosen so that the atomic mass is 1, Planck's
constant is 1, and the spatial period of the kicks is $2\pi$. The
dimensionless parameters $k,\tau,\eta$ are expressed in terms of the
physical parameters as follows: $k=\kappa/\hbar$, $\tau=\hbar
TG^2/M$, $\eta=MgT/(\hbar G)$, where $M,T,\kappa$ are the atomic
mass, the kicking period, the kick strength, and $2\pi/G$ is the
spatial   period of the kicks.  $\hat X$ is the position operator
(along the vertical direction) and the kicking potential $V({\hat
X})=\cos({\hat X})$ in experiments. Hamiltonian (\ref{ham1}) is
written in a special, time dependent gauge \cite{FGR03}, in which
the canonical momentum operator is given by ${\hat P}+\eta t/\tau$.
This choice of a gauge makes (\ref{ham1}) invariant under spatial
translations by $2\pi$, so, by Bloch theory, the quasi-momentum
$\beta$ is conserved. With the present units,  $\beta$ is  the
fractional part of $\hat P$. The dynamics at fixed $\beta$ are
formally those of a rotor with angular coordinate
$\theta=X$mod$(2\pi)$. Let $|\psi_n\rangle$ denote the state 
of the rotor immediately after the $n$-th kick; then
$|\psi_{n+1}\rangle={\hat U}_n|\psi_n\rangle$, where the unitary
operators ${\hat U}_n$ are given, in the $\theta$-representation,
by:
\begin{equation}
\label{kr}
 {\hat U}_n\;=\;e^{-ik V({\theta})}\;
e^{-i\frac{\tau}{2}(-i\partial_{\theta}+\beta+\eta/2+\eta n)^2}\;.
\end{equation}
For $\eta=0$, ${\hat U}_n$ does not depend on $n$, and coincides
with the propagator of the generalized Kicked Rotor.
Multiplication of wavefunctions $\psi(\theta)$ by $\exp(i m\theta)$,
($m\in\ZM$) generates the discrete unitary group of (angular)
momentum translations. For special values of $\tau$ and $\beta$  a
nontrivial subgroup of such translations commutes with the KR
propagator. This leads to
 a special dynamical behaviour, called KR-resonance \cite{IS80}.
 We define  the
order of a KR resonance as the minimum index of a commuting
subgroup; or, the least positive integer $\ell$ such that (\ref{kr})
commutes with multiplication by $\exp(i\ell\theta)$.  KR resonances
occur if, and only if \cite{refwhen}, $\tau$ is commensurate to $2\pi$, and the
quasi-momentum $\beta$ is rational. Indeed, momentum
translations by multiples of an integer $\ell$ leave (\ref{kr})
invariant if, and only if, (i) $\tau=2\pi p/q$ with $p,q$ coprime
integers, (ii) $\ell=mq$ for some integer $m$, and (iii) $\beta={\nu}/mp+mq/2$
mod$(1)$, with $\nu$ an arbitrary integer.
 In the
following we restrict to "primary" resonances, which have $m=1$ and
$\ell=q$, and generically denote $\beta_r$ the resonant values of
quasi-momentum. The KR propagator at exact resonance
is obtained
on substituting $\tau=2\pi p/q$, $\beta=\beta_r$, and $\eta=0$ in (\ref{kr}). Using
Poisson's summation formula, it may be written in the form:
\begin{equation}
\label{freerot}
 {\hat U}_{\mbox{\rm\tiny res}}\psi(\theta)\;=\;
 e^{-ik V(\theta)}
\sum\limits_{s=0}^{q-1}
 G_s\;{\psi}(\theta-2\pi s/q)\;,
 \end{equation}
where
\begin{equation}
\label{gauss} G_s\equiv G_s(p,q,\beta_r)\;=\;\frac{1}{
q}\sum\limits_{l=0}^{q-1} e^{-\pi i p(l+\beta_r)^2/q}\;e^{2\pi i
sl/q}\;,
 \end{equation}
so that $|G_s|=q^{-1/2}$. Now let $\tau=2p\pi/q+\epsilon$, $\beta=\beta_r+\delta\beta$; and denote
$\phi_n=\delta\beta+\eta/2+\eta n$. We may write
\begin{equation}
\label{right} {\hat U}_n\;=\; {\hat U}_{\mbox{\rm\tiny res}}
\exp(-i\frac{\epsilon}{2}(-i\partial_{\theta}+\beta_r)^2)\;\exp
(-\tau\phi_n\partial_{\theta})\;.
\end{equation}
 Here, and in the following, phase factors only dependent on $\beta$
and $n$ are disregarded. Thanks to
eqn.(\ref{freerot}), eqn.(\ref{right}) may be rewritten in the
following form :
\begin{equation}
\label{rright} ({\hat U}_n\psi)(\theta)\;=\; e^{-ik
V(\theta)}\sum\limits_{s=0}^{q-1}\;G_s \;{\tilde\psi}(\theta-2\pi
s/q-\tau\phi_n)\;,
\end{equation}
where:
\begin{equation}
\label{prop} {\tilde\psi}(\theta)\;=\;
e^{-i\frac{\epsilon}{2}(-i\partial_{\theta}+\beta_r)^2}\psi(\theta)\;.
\end{equation}
\begin{figure}
  \includegraphics[width=8cm,angle=0]{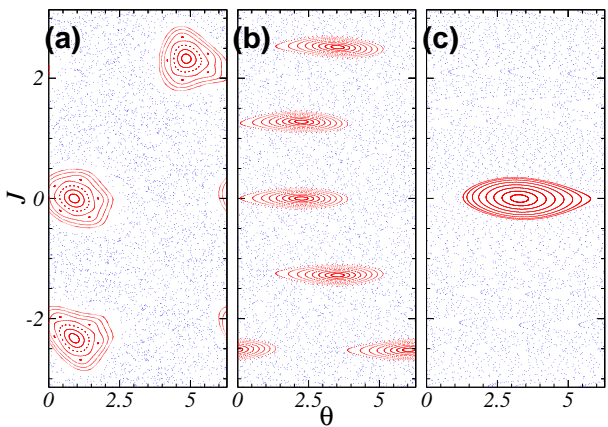}
  \caption{(Color online) Phase portraits of maps ${\cal F}^{(T)}_0$  on the 2-torus, for
  (a)  $T=2, \delta _s =(-1)^{s+1}\pi, \tilde k= -0.395, \tau\eta=1.122,
  ({\mathfrak p},{\mathfrak j})=(3,1)$ $(\tau/2\pi =0.475,
  \epsilon =-0.157)$; (b) $T=1, \delta _s =0, \tilde k= 0.032, \tau\eta=1.253,
  ({\mathfrak p},{\mathfrak j})=(5,1)$ $(\tau/2\pi =0.502,
  \epsilon =0.013)$ and (c) $T=1, \delta _s =20\pi /13, \tilde k= 0.040, \tau\eta=1.455,
  ({\mathfrak p},{\mathfrak j})=(1,1)$ $(\tau/2\pi =0.541,
  \epsilon =0.016)$.}
\label{phsp}
\end{figure}
If $\epsilon$ is granted the formal role  of  Planck's constant, then operator
(\ref{prop}) has the form
of a unitary propagator for a generalized free rotor \cite{SH}, so
quasi-classical methods may be used to investigate the
small-$\epsilon$ regime. We define the
$\epsilon$-classical momentum operator ${\hat
I}=-i\epsilon\partial/\partial\theta$ \footnote{the momentum $I$ is
opposite in sign to physical momentum whenever $\epsilon<0$. This
convention is different from the one which was used in
ref. \cite{FGR03} and allows for simpler notations, at the cost of accepting negative values of the ''Planck's constant'' $\epsilon$.}.
Denoting ${\tilde k}\equiv k\epsilon$, the $\epsilon$-quasiclassical
asymptotic regime is defined by $\epsilon\to 0$ at constant ${\tilde
k},I$. Using the explicit form of the integral kernel for
(\ref{prop}) \cite{SH}, the transition amplitude from
$\theta=\theta_0$ at time $0$ to $\theta=\theta_n$ after  $n$ kicks
 is given by:
\begin{gather}
\langle \theta_n|{\hat U}_{n-1}\ldots{\hat U}_0|\theta_0\rangle= (2\pi
i\epsilon)^{-n/2} \sum\limits_{({\bf m},{\bf s})\in\Omega_n}
G_{s_0}\ldots G_{s_{n-1}}\times\nonumber\\
\times\int_{0}^{2\pi}\ldots\int_{0}^{2\pi}d\theta_1\ldots
d\theta_{n-1} e^{i\epsilon^{-1}S_{{\bf m},{\bf s}}(\theta_0,\theta_1,\ldots,\theta_n)}\;, \label{trans}\end{gather}
where ${\bf m}$ and ${\bf s}$ are vectors with $(n-1)$ integer components, $\Omega_n
\equiv\ZM^{n-1}\times\{1,\ldots,q\}^{n-1}$, and
\begin{gather}
 S_{{\bf m},{\bf s}}(\theta_0,\ldots,\theta_n)=
\sum\limits_{t=1}^{n} \{-{\tilde k}
V(\theta_t)+\nonumber\\
+\frac12(\theta_t-\theta_{t-1}-2\pi s_t/q-2m_t\pi-\tau\phi_t)^2\}\;.
\label{action}\end{gather}
 Replacing (\ref{action}) in
(\ref{trans}), and using the stationary phase approximation in
individual terms in the sum on the rhs of (\ref{trans}), we find
that, at small $|\epsilon|$, (\ref{prop}) propagates
along rays,  which satisfy  the equations :
\begin{eqnarray}
\label{map} \theta_{t+1}\;&=&\; \theta_{t}\;+\;I_t\;+\;
\tau\phi_t\;+2\pi
s_t/q\;\mbox{\rm mod}\;2\pi\nonumber\\
I_{t+1}\;&=&\;I_{t}\;-\;{\tilde k}V'(\theta_{t+1})\;,
\end{eqnarray}
or, defining $J_t\equiv I_t+\tau\phi_t+2\pi s_t/q$, and
$\delta_t\equiv 2\pi(s_{t+1}-s_t)/q$,
\begin{eqnarray}
\label{map1} J_{t+1}\;&=&\;J_{t}\;+\delta_t +\tau\eta - {\tilde
k}\;V'(\theta_{t+1})\;,
\nonumber\\
\theta_{t+1}\;&=&\; \theta_{t}\;+ J_t\;\mbox{\rm mod}(2\pi)\;.
\end{eqnarray}
For each value of $t$, (\ref{map1}) defines a map ${\cal F}_t$ on the cylinder;
however,  since the choice of the integers $s_1,s_2,\ldots$ is totally
arbitrary whenever $q>1$, such maps  do not, in general, uniquely
define  a classical dynamical system.  The $s_t$ may be removed by
changing variables to $\vartheta=q\theta$, but this calls into play
the function $V'(\vartheta/q)$, which is not a single-valued
function in $\vartheta\in[0,2\pi]$, except in the case when $V(\theta)$ is a $2\pi/q$-periodic
function; then eqs. (\ref{map1}) reduce to a single map,
and the theory proceeds essentially identical as in  the case $q=1$.
In all other cases,  exponentially many different maps
enter the game upon
iterating eqs.(\ref{map1}), and so no $\epsilon$-classical limit
proper  exists. In spite of that, we shall presently show how a stability requirement
singles out special families of  rays, which give distinguished contributions in
the dynamics, ultimately resulting in QAM.
In stationary phase approximation, each ray
(\ref{map}) contributes a term $q^{-n/2}|$det$({\mathfrak M})|^{-1/2}
\exp(iS_{{\bf s},{\bf m}}/\epsilon+i\Phi_{\bf s})$
in (\ref{trans}), where $S_{{\bf s},{\bf m}}$ is the action (\ref{action}) computed
along the given ray, $\Phi_{\bf s}$ collects phases from
the $G_{s_t}$ and from Maslov indices,
and $\mathfrak M$ is the matrix of  2nd derivatives of
(\ref{action}) with respect to the angles
$\theta_1,\ldots,\theta_{n-1}$.
Stability of a ray is related to the behavior
of the prefactor $|$det$({\mathfrak M})|^{-1/2}$ as a  function of "time" $n$.
$\mathfrak M$ is a tridiagonal Jacobi
matrix, with off-diagonal elements equal to $-1$, and diagonal
elements given by $-{\tilde k}V''(\theta_t)+2$, where $\theta_t$ are
the angles along the ray.
For a large number $n$ of kicks, most
choices of ${\bf s}\in\{1,\ldots,q\}^n$ are essentially random. The
same may be assumed to be true of the diagonal elements of $\mathfrak M$, and so
$\mathfrak M$ has a positive Lyapunov exponent, due to Anderson
localization. It follows that $|$det$({\mathfrak M})|$ exponentially increases with $n$ (as may be seen, e.g.,
 from the Herbert-Jones-Thouless formula \cite{PF}). Therefore, such rays carry
exponentially small contributions, and  their  global effect is determined by interference of exponentially
many such contributions.
In contrast, distinguished  contributions are given
by those rays, whose matrices $\mathfrak M$ have
extended states, thanks to absence of diagonal disorder. The simplest such case
occurs when the diagonal elements of $\mathfrak M$ are a periodic sequence.
This in particular happens when  $\delta_t$ is a periodic sequence,
and rays are in such cases  related to stable periodic orbits of certain
classical dynamical systems, which are constructed as follows.
Let $\delta_{t+T}=\delta_t$ for some $T$
and all $t$. Then map ${\cal F}_t$  (\ref{map1}) periodically depends on
"time" $t$; so, for each choice  of  $t'$ with $0\leq t'<T-1$, one may introduce
a "map over one period"
${\cal F}^{(T)}_{t'}\equiv{\cal F}_{t'+T-1}\circ\ldots\circ{\cal F}_{t'}$, whose iteration
determines rays (\ref{map1}) at every $T$-th kick after the $t'$-th one.
As this  map
is $2\pi$-periodic in $J,\theta$, it defines a
dynamical system on the 2-torus. Systems that way constructed with different $t'$ are obviously
conjugate to each other,
so the periodic orbits of any of them  one-to-one correspond
to the periodic orbits that are obtained for  $t'=0$.
As a result,  to each periodic orbit of ${\cal F}^{(T)}_0$ (on the 2-torus)
a ray (\ref{map}) is associated, which is periodic in position space; therefore,
its matrix $\mathfrak M$ has periodic diagonal elements.
If the orbit has period $\mathfrak p$, then
 the  corresponding ray (\ref{map1}) satisfies
$J_{(l+{\mathfrak p})T}=J_{lT}+2\pi{\mathfrak j}$ for all integer
$l$, where $\mathfrak j$ is the "jumping index" of the periodic orbit. This is
equivalent to $I_{(l+{\mathfrak p})T}=I_{lT}-2\pi (s_{(l+{\mathfrak p})T}-s_{lT})/q-\tau\eta T{\mathfrak
p}+2\pi{\mathfrak j}$ and so, along such a ray,  the physical
momentum $I/\epsilon$ linearly increases (or decreases) with average
acceleration
\begin{equation}
\label{acce}
a=\epsilon^{-1}\left\{2\pi{\mathfrak j}({\mathfrak p}T)^{-1}-\Delta_T-\tau\eta\right\}\;,
\end{equation}
where $\Delta _T =T^{-1}\sum _{s=0}^{T-1}\delta _s$.
Finally, stability of such rays, as determined by the behavior of det$({\mathfrak M})$ as a function
of "time" $n$, is controlled by the Lyapunov exponent, and so is equivalent to  dynamical stability
of the corresponding periodic orbits
 \footnote{  The Lyapunov
exponent of the tridiagonal Jacobi matrix $\mathfrak M$ is decided by products of $2\times 2$ transfer matrices; stability
of the periodic orbit associated to $\mathfrak M$ is decided by products of tangent maps of the maps ${\cal F}_t$.  Direct calculation shows that transfer matrices and tangent maps are related
by a constant similarity transform, so the two types of products have similar behaviors.}.\\ 
  In summary: whenever
$V(\theta)$ is not $2\pi/q$-periodic,  no $\epsilon$-classical limit
exists for the dynamics (\ref{trans});  QAM may nevertheless exist,
associated with stable "accelerator rays", that are
associated with the stable periodic orbits of a family of maps of
the 2-torus. There is one such map for each choice of a periodic
sequence in $\{1,\ldots,q\}^{\NM}$. The simplest choice is
$\delta_t=0$ ; the relevant map (\ref{map1}), and the
acceleration formula (\ref{acce}), are then the same as in the case
$q=1$ \cite{FGR03}. In Fig. \ref{pac} we show numerical evidence for
QAM associated with  the resonances at $q=2$, $p=1$ ((a) and (b)) and
at $q=13$, $p=7$ (c).
Here, the kicking potential is
$V(\theta)=\cos(\theta)$. For the given parameter values,
three QAM are clearly detected: two around the $q=2$, $p=1$ resonance, one
near the $q=13$, $p=7$ one. They correspond, via eq.
(\ref{acce}), to stable periodic orbits of maps ${\cal F}^{(T)}_0$ with $T=1$
and with $T=2$. The stable islands of these orbits are shown in Fig. \ref{phsp}. \\
The present theory suggests an unsuspected richness of QAM, associated
with the dense set of higher-order resonances. If produced with ideal, infinite resolution, figures in the style of  Fig. \ref{pac} might reveal that
QAM are essentially ubiquitous; however, some QAM associated with resonances of low order $q>1$
should be observable  already on the present level of experimental resolution. Our numerical
simulations have exposed a fine texture of seemingly QAM-like structures; on the available level
of precision, however, most of them are so vague, that it is impossible to decide to which resonance
they belong. Those for which this question could be answered were in all cases found to correspond to some stable orbits, in agreement with the above theory.
 On the other hand, for a few  of the periodic orbits we have computed,
no partner QAM  could be detected. This may be due to the fact that, at given parameter values, many different orbits coexist, which are related to different resonances, hence to different values of the pseudo-Planck constant $\epsilon$. The hierarchical rules that determine their relative  "visibility"  are not known at this stage. In general, one may expect stronger QAM near lower order resonances, yet exceptions are not rare, see  Fig.1 (c).

We  thank G. Summy for communicating results obtained by his group,
prior to publication, and S. Fishman for his constant attention and
precious comments in the course of this work.

\end{document}